\documentclass[12pt]{iopart}

\makeatletter
\expandafter\let\csname equation*\endcsname\relax
\expandafter\let\csname endequation*\endcsname\relax
\makeatother

\usepackage{amsmath,amssymb,mathtools}
\usepackage{bm}
\usepackage{xcolor}
\usepackage{graphicx}
\usepackage{tikz}
\usetikzlibrary{decorations.pathmorphing,decorations.pathreplacing,arrows.meta,calc}

\newcommand{\Sc}{\mathcal{S}}
\newcommand{\N}{\mathcal{N}}
\newcommand{\dd}{\mathrm{d}}
\newcommand{\lp}{\ell_{\mathrm{P}}}
\newcommand{\half}{\tfrac{1}{2}}

\begin{document}

\title{On a geometric interpretation of the $1/4$ factor in black hole entropy}

\author{Ira Wolfson}

\address{Department of Electrical and Electronic Engineering,
Braude Academic College of Engineering, Karmiel 2161002, Israel}

\ead{wolfsoni@braude.ac.il}

\begin{abstract}
The Bekenstein--Hawking entropy $S = k_B A/4\lp^2$ contains a coefficient
$1/4$ that has been calculated by many approaches to quantum gravity.
We show that this factor arises from the causal geometry of
any null boundary in a four-dimensional Lorentzian spacetime.
A heuristic accessibility argument is supported by a purely
geometric derivation using the canonical symplectic structure
on the cotangent bundle and the causal splitting of the space
of null rays.
No gravitational field equations are invoked.
\end{abstract}

\maketitle

\section{Introduction}

The entropy of a black hole is
\begin{equation}
  S = \frac{k_B A}{4\,\lp^2},
  \qquad
  \lp^2 = \frac{G\hbar}{c^3},
\end{equation}
where for the remainder of this paper $k_B=1$, and the other physical constants--($\hbar,G,c$)--are kept explicitly but appear only through the Planck-scale $\ell_p$.
The universality of the coefficient $1/4$ suggests a
geometric origin independent of microscopic details,
as it arises in diverse frameworks including semiclassical
gravity, Noether-charge methods, Euclidean path integrals,
thermodynamic arguments, and combinatorial approaches
\cite{Bekenstein1973,Hawking1975,Wald1993,IyerWald1994,
GibbonsHawking1977,Jacobson1995,Davidson2019}.

We argue that this factor is the fraction of phase space
accessible across any null causal boundary in a four-dimensional
Lorentzian spacetime.
We do not attempt a microscopic derivation of black hole entropy;
instead, we identify a kinematic mechanism in null causal geometry
that produces the factor $1/4$ independently of any particular
dynamics or UV completion.

\section{Heuristic accessibility argument}

We begin with a heuristic overview of the reduction;
the formal derivation follows in Section~3.

A point in the phase space of a four-dimensional
Lorentzian spacetime is specified by four coordinates
and four conjugate momenta---eight dimensions in total.

\medskip
\noindent\textbf{(1) Null mass-shell constraint\footnote{In reparametrization-invariant language,
$Q=0$ acts as a Hamiltonian constraint generating the geodesic flow.} ($8\to 7\to 6$).}
Imposing a mass-shell or null condition
$Q = g^{ab}p_a p_b = 0$ makes one momentum a function of the
others, and the resulting evolution sweeps out orbits
along the constraint surface, removing the conjugate
coordinate as an independent degree of freedom.
Two dimensions are lost; six remain (in the GR formalism, this is given in \cite{DonnellyFreidel2016}).

\medskip
\noindent\textbf{(2) Null surface ($6\to 4$).}
Confinement to a null hypersurface $\N$ in the base
manifold fixes the transverse coordinate, removing one
degree of freedom.
In Lorentzian signature, the normal to a null surface is
also tangent to it---a feature with no Riemannian
analogue---so the generator direction within $\N$ is
not independent but already determined,
removing a second degree of freedom.
What remains is a two-dimensional spacelike
cross-section $\Sc$, carrying two canonical pairs (recovered in GR in \cite{DonnellyFreidel2016,HopfmullerFreidel2017}).

\medskip
\noindent\textbf{(3) Conscripted time ($4\to 2$).}
The two canonical pairs on $\Sc$ include directions tangent
to the generators of $\N$.
Deformations along these directions slide points along
the same null rays without changing the surface;
they carry no independent information.
One coordinate is therefore conscripted to track position
along the generators: the induced metric on $\Sc$ is positive
definite, so no geometric evolution parameter exists,
and one pair must serve as a relational clock for the
other~\cite{Rovelli2002,GambiniPullin2011,RovelliVidotto2015}.
In the language of the method of characteristics, the time
coordinate drops out---$F(q_1,p_1,q_2,p_2,t)\to
F(q_1,p_1,q_2,p_2)=C$---and $q_2$ inherits the role of
classical time, parametrizing the family of solutions while
$p_2$ selects among them.
One pair is consumed; one survives---the equivalent of
``dressing time'' in \cite{CiambelliFreidelLeigh2024}.

\medskip
\noindent\textbf{(4) Causal restriction ($2\to 1$ observable).}
A null hypersurface is a one-way causal boundary.
After the clock conscription, only one canonical pair survives.
This pair carries both future-directed and past-directed data,
but an observer confined to one side receives signals from only
one of these two components.
Only half the pair is accessible (recovered in GR in
\cite{WaldGR1984}).

\medskip
The accessible fraction of the naive surface phase space is thus
\begin{equation}
  \frac{\half\;\text{pair}}{2\;\text{pairs}} = \frac{1}{4}\,.
\end{equation}

\noindent\textit{Illustration.}\quad
Schwarzschild coordinates are singular at the horizon
and therefore unsuitable for a rigorous treatment
(Section~5 uses regular Eddington--Finkelstein coordinates),
but they make the dimensional reduction vivid.
In these coordinates,
$\dd s^2 = -f\,\dd t^2 + f^{-1}\dd r^2 + r^2\dd\Omega^2$
with $f=1-r_s/r$.
The null condition $Q=0$ constrains the momenta.
The null surface sits at $r=r_s$, where $f=0$.
There $\dd s^2=0$ implies that $t\to t+\epsilon$ for
arbitrary $\epsilon$ leaves the geometry unchanged:
the time coordinate loses all local meaning.
Steps~(1) and~(2) have removed $(t,p_t)$ and $(r,p_r)$;
what remains is the Riemannian 2-sphere
$(\theta,\varphi,p_\theta,p_\varphi)$---two canonical pairs.
Step~(3) conscripts one angular coordinate as the
evolution parameter, leaving a single pair.
Step~(4) halves it.

This argument is heuristic; a purely geometric formal realization
is given next.

\section{Formal geometric derivation}

Throughout this section $(M,g)$ is a four-dimensional
time-oriented Lorentzian manifold\footnote{For any $n$-dimensional manifold $M$, the cotangent
bundle $T^*M$ has dimension $2n$. In our case $\dim M = 4$,
so $\dim T^*M = 8$.}, so that $\dim T^*M = 8$.

The reduction begins from the full cotangent bundle
$T^*M$ rather than from the null cone directly for self-containment and clarity.
Each step is then geometrically explicit, and the
origin of every factor of 2 is transparent.

The canonical one-form on $T^*M$ and the null constraint
$Q = g^{ab}p_a p_b$ involve different types of contraction:
the former uses only the natural dual pairing between
vectors and covectors, while the latter requires the
spacetime metric.
This distinction is fundamental: the symplectic structure
is a property of the cotangent bundle alone, requiring no
metric, while the null condition is a statement about the
spacetime geometry. These two structures are logically
independent \cite{AbrahamMarsden1978,Arnold1989}.
Throughout, indices are raised and lowered using $g_{ab}$ only
when explicitly indicated.

We now formalize the heuristic argument using the symplectic
geometry of null covectors.

\subsection{Canonical phase space}

On $T^*M$ define the canonical one-form and symplectic
form\footnote{We follow the sign convention of
\cite{Arnold1989}, with $\omega=-\dd\theta$.}
\begin{equation}
  \theta = p_a\,\dd x^a,
  \qquad
  \omega = -\dd\theta = \dd x^a\wedge\dd p_a\,.
\end{equation}

Define the quadratic function on $T^*M$
\begin{equation}
Q(x,p)=g^{ab}(x)\,p_a\,p_b\,.
\end{equation}
The condition $Q=0$
is not a dynamical mass-shell constraint but a geometric condition
selecting the null covectors in $T^*M$, determined entirely by
the Lorentzian metric.

The null covectors form the hypersurface
\begin{equation}
\mathcal{C}=\{Q=0\}\subset T^*M\,.
\end{equation}
The symplectic structure $\omega$ associates to $Q$ a vector
field $X_Q$---the Hamiltonian vector field of $Q$---on
$T^*M$ via $\iota_{X_Q}\omega = \dd Q$.
On $\mathcal{C}$ the integral curves of $X_Q$ satisfy
\begin{equation}\label{eq:hamilton}
\frac{\dd x^a}{\dd\lambda}
  = \frac{\partial Q}{\partial p_a}
  = 2\,g^{ab}\,p_b\,,
\end{equation}
where $\lambda$ is the curve parameter (not a time coordinate).
The tangent vector is proportional to the covector $p_a$
with index raised by the metric.
On $\mathcal{C}$ this vector is null; the integral curves
are null geodesics.

\subsection{Space of null rays}

Since $Q$ is constant on $\mathcal{C}$, any vector $V$
tangent to $\mathcal{C}$ satisfies $\dd Q(V)=0$.
But $\iota_{X_Q}\omega=\dd Q$ by definition, so
\begin{equation}
\omega(X_Q,\,V)=0
\qquad\text{for all } V \in T\mathcal{C}\,.
\end{equation}
That is, $X_Q$ lies in the kernel of $\omega|_{\mathcal{C}}$.

Moreover, $X_Q$ is tangent to $\mathcal{C}$: the flow
preserves $Q$ because $X_Q(Q)=\omega(X_Q,X_Q)=0$ by
antisymmetry.

Dimension counting:
\begin{align}
\dim T^*M &= 8\,,\\
Q=0 &\Rightarrow \dim\mathcal{C}=7\,,\\
\text{quotient by }X_Q &\Rightarrow 6\,.
\end{align}
Quotienting $\mathcal{C}$ by the one-dimensional orbits of
$X_Q$ yields the space $\mathcal{R}$ of unparametrized null
geodesics (null rays).
The restriction $\omega|_{\mathcal{C}}$ has a one-dimensional
kernel spanned by $X_Q$; it therefore descends to a
non-degenerate closed 2-form on the 6-dimensional quotient
$\mathcal{R}$, making $\mathcal{R}$ a symplectic manifold
\cite{DonnellyFreidel2016}.

\subsection{Restriction to a cross-section}

Let $\N$ be a null hypersurface and $\Sc\subset\N$ a closed spacelike
2-surface that serves as a cross-section of $\N$.
Each point of $\Sc$ determines a unique generator of $\N$,
which is a single point in $\mathcal{R}$.
The generators through $\Sc$ therefore define a 2-dimensional
submanifold $G\subset\mathcal{R}$.

The restriction $\omega|_G$ vanishes:
deformations within the family of generators of a null
hypersurface carry no symplectic flux
\cite{DonnellyFreidel2016,HopfmullerFreidel2017,Speranza2018}.
Hence $G$ is isotropic\footnote{A submanifold $G$ is called \emph{isotropic}
by analogy with the metric case: just as a null vector $v$
satisfies $g(v,v)=0$, the tangent vectors of $G$ satisfy
$\omega(u,v)=0$ for all $u,v\in TG$---they are isotropic with
respect to the symplectic form. This is a symplectic condition,
distinct from but motivated by the metric notion of isotropy.}.

The symplectic complement $TG^\perp$---the set of all
directions in $T\mathcal{R}$ that have vanishing symplectic
product with every vector tangent to $G$---has dimension
$6-2=4$, i.e.\ two canonical pairs.
These encode all phase-space data in $\mathcal{R}$
coupled to the surface and constitute the naive
phase-space content of a surface element.

Because $G$ is isotropic, $TG\subset TG^\perp$.
Quotienting by $TG$ removes the two directions tangent to
$G$, leaving
\begin{equation}
\dim\!\left(TG^\perp/TG\right) = 4-2 = 2\,,
\end{equation}
i.e.\ one canonical pair per surface element.
This pair corresponds to the expansion--shear conjugate
structure identified in
\cite{HopfmullerFreidel2017,ChandrasekaranSperanza2021}.
The quotient $TG^\perp/TG$ is the kinematic analogue of
the \emph{dressing time} construction of
Ciambelli, Freidel, and Leigh \cite{CiambelliFreidelLeigh2024}.
The reduction is mandatory: the induced metric on
$\Sc$ is positive definite, so Hamilton's equations on
$T^*\Sc$ admit no geometric evolution parameter.
In Rovelli's language \cite{Rovelli2002}, the two
canonical pairs are partial observables; only their
correlations are complete observables, so one pair
must serve as a relational clock for the other.
Entropy counts independent phase-space labels; a pair
serving as a relational clock is not an independent
label of the state but a parametrization of correlations,
and therefore does not contribute to the microstate count.
This mechanism is independent of the Raychaudhuri
equation and of any gravitational dynamics \cite{Raychaudhuri1955}.

\subsection{Causal halving}

Time orientation splits $\mathcal{C}$ into future-pointing
($\mathcal{C}^+$) and past-pointing ($\mathcal{C}^-$) null covectors.
Since $Q(x,-p)=Q(x,p)$, the map $(x,p)\mapsto(x,-p)$
preserves $\mathcal{C}$ and maps orbits of $X_Q$ to orbits,
so it descends to an involution on $\mathcal{R}$ that
exchanges the future-pointing and past-pointing components
$\mathcal{R}^\pm$.

Under $p_a\mapsto -p_a$ the symplectic form transforms as
$\omega\mapsto -\omega$, so the reduced symplectic form on
$\mathcal{R}$ also reverses sign.
The Liouville measure \cite{Arnold1989}, being proportional to
$|\omega_{\mathcal{R}}^{\,3}|$, is invariant:
\begin{equation}
  \mu(\mathcal{R}^+) = \mu(\mathcal{R}^-)
  = \half\,\mu(\mathcal{R})\,.
\end{equation}

This splitting is dynamically empty. The flow of
$X_Q$ is time-reversal symmetric not by virtue of being a
Hamiltonian flow in general, but because $Q$ is a geometric
constraint on the cotangent bundle whose quadratic structure
in $p$ gives $Q(x,-p) = g^{ab}(-p_a)(-p_b) = Q(x,p)$. The
map $(x,p)\mapsto(x,-p)$ therefore preserves $\mathcal{C}$
and maps orbits of $X_Q$ to orbits. The full phase space is
causally accessible to a freely falling observer who
crosses $\N$.
A null hypersurface is the only geometric structure in Lorentzian spacetime that locally partitions
a neighborhood into causally disconnected regions across which signals propagate in only one direction.
As such, it defines a causal boundary.
It divides a neighborhood of $N$ into two regions:
future-directed null rays from $S$ propagate into one region,
while past-directed rays propagate into the other.
An observer confined to one side receives signals from only one type of rays.

The causal structure of $(M,g)$---encoded in the
distinction between the causal future $J^+(\N)$ and
causal past $J^-(\N)$---therefore restricts the
\emph{observable} portion of the phase space to one
temporal component.
At each surface element, the single canonical pair
identified in Section~3.3 carries both
$\mathcal{R}^+$ and $\mathcal{R}^-$ data, but only
one component is accessible:
\begin{equation}
  \mu_{\text{obs}} = \half\,\mu_{\text{full}}\,.
\end{equation}

The full phase space remains intact.
The halving is not a dynamical reduction but a causal
restriction on observational access, determined entirely
by the Lorentzian causal structure of $\N$.

\section{Entropy}

We now collect the reduction chain.
Starting from eight phase-space dimensions per spacetime
point on $T^*M$:
the null constraint $Q=0$ removes one (Section~3.1);
the quotient by $X_Q$ removes another (Section~3.2);
the symplectic complement $TG^\perp$ selects the four
dimensions coupled to the surface---two canonical
pairs---which constitute the naive phase-space content of
a surface element (Section~3.3);
the isotropic quotient $TG^\perp/TG$ removes the remaining
two, leaving one canonical pair.

Time orientation (Section~3.4) divides the surviving
pair into future and past components of equal measure.
The full phase space remains intact, but a causal boundary
permits observation of only one temporal direction.
An observer on one side therefore has access to half of the
surviving pair.

According to common wisdom, entropy counts microstates consistent
with the observed macrostate. For an exterior observer, the
macrostate is defined by causally accessible data alone; microstates
in $\mathcal{R}^-$ are causally inaccessible and therefore logically
excluded from the count. No coarse-graining or tracing-out is
required: the exclusion follows directly from the definition of
the macrostate and halves the accessible count.

The observable fraction of the naive surface phase space is
\begin{equation}
\frac{\half\;\text{pair}}{2\;\text{pairs}}=\frac{1}{4}\,.
\end{equation}

Whatever Planck-area cell count $N = A/\lp^2$ the UV
completion provides---whether from combinatorial
tiling~\cite{Davidson2019} or any other
microscopic framework---the combinatorics supplies
the count while the geometry fixes the accessible
fraction, yielding
\begin{equation}
S=\frac{A}{4\,\lp^2}\,.
\end{equation}

\section{Application to black holes}

\begin{figure}[t]
\centering
\resizebox{\textwidth}{!}{%
\begin{tikzpicture}[scale=1.1, >=Stealth,
  every node/.style={font=\small}]

\coordinate (bif) at (0,0);
\coordinate (iplus) at (1.8,1.8);
\coordinate (iminus) at (1.8,-1.8);
\coordinate (i0) at (3.6,0);
\coordinate (singfut) at (-1.8,1.8);
\coordinate (singpast) at (-1.8,-1.8);
\coordinate (i0left) at (-3.6,0);

\fill[blue!5] (bif) -- (iplus) -- (i0) -- (iminus) -- cycle;
\fill[red!5] (bif) -- (singfut) -- (iplus) -- cycle;

\draw[thick, decorate,
  decoration={zigzag, segment length=4pt, amplitude=1.5pt}]
  (singfut) -- (iplus);
\node[above=4pt, font=\scriptsize] at (0,1.8) {singularity};
\draw[thick, decorate,
  decoration={zigzag, segment length=4pt, amplitude=1.5pt}]
  (singpast) -- (iminus);

\draw[thick] (iplus) -- (i0) -- (iminus);
\draw[thick] (singfut) -- (i0left) -- (singpast);
\draw[thick, dashed, gray!60] (bif) -- (iminus);
\draw[thick, dashed, gray!60] (bif) -- (singpast);
\draw[thick, dashed, gray!60] (bif) -- (i0left);

\draw[very thick, blue!70!black] (bif) -- (iplus);
\node[blue!70!black, above left=1pt, font=\footnotesize\itshape]
  at (0.5,0.5) {$\mathcal{N}$};

\coordinate (S) at (0.8,0.8);
\fill[black] (S) circle (1.8pt);
\node[below right=1pt, font=\footnotesize\bfseries] at (S)
  {$\mathcal{S}$};

\draw[->, thick, blue!70!black] (0.2,0.2) -- (0.5,0.5);
\draw[->, thick, blue!70!black] (1.0,1.0) -- (1.35,1.35);
\node[blue!70!black, font=\tiny, rotate=45, above=1pt]
  at (1.18,1.18) {generators};

\draw[->, thick, green!50!black, densely dashed]
  (S) -- ++(0.9,-0.9);
\draw[->, thick, green!50!black, densely dashed]
  (S) -- ++(1.3,-1.3);
\node[green!50!black, font=\tiny, align=center]
  at (2.5,-0.9) {outgoing\\[-1pt](accessible)};

\draw[->, thick, red!70!black, densely dashed]
  (S) -- ++(-0.65,0.65);
\draw[->, thick, red!70!black, densely dashed]
  (S) -- ++(-0.9,0.9);
\node[red!70!black, font=\tiny, align=center]
  at (-1.3,0.95) {ingoing\\[-1pt](inaccessible)};

\coordinate (obs) at (2.3,0.3);
\fill[green!30!black] (obs) circle (1.2pt);
\draw[thick] (obs) circle (2.5pt);
\node[above right=2pt, font=\tiny] at (obs) {observer};

\node[blue!30!black, font=\tiny\itshape] at (1.6,0.6) {exterior};
\node[red!30!black, font=\tiny\itshape] at (-0.5,1.25) {interior};
\node[below right=1pt, font=\tiny] at (i0) {$i^0$};

\node[anchor=north west, font=\tiny, align=left]
  at (4.0, 1.7) {%
  \textbf{Phase-space reduction:}\\[3pt]
  $T^*\!M$: \textbf{8} dimensions\\[1pt]
  $Q=0$: $8\to\mathbf{7}$\\[1pt]
  $\mathcal{R}=\mathcal{C}/X_Q$: $7\to\mathbf{6}$\\[1pt]
  $TG^\perp$: $6\to\mathbf{4}$ (2 pairs)\\[1pt]
  $TG^\perp\!/TG$: $4\to\mathbf{2}$ (1 pair)\\[1pt]
  causal halving: $2\to\mathbf{1}$ obs.};

\draw[->, thick, gray!50] (3.9, 0.0) -- (3.0, 0.0);

\node[draw, thick, rounded corners=2pt,
  fill=yellow!12, inner sep=4pt,
  font=\small]
  at (5.8, -1.0) {%
  $\displaystyle\frac{\tfrac{1}{2}\;\text{pair}}
  {2\;\text{pairs}}=\frac{1}{4}$};

\end{tikzpicture}%
}
\caption{Penrose diagram of a Schwarzschild black hole
with the phase-space reduction chain.
The cross-section $\Sc$ on the future horizon $\N$
is traversed by null rays in two temporal directions.
Outgoing rays (green, dashed) reach the exterior observer;
ingoing rays (red, dashed) fall into the interior and are
causally inaccessible.
The reduction chain (right) traces the eight cotangent-bundle
dimensions to a single observable degree of freedom,
yielding the factor $1/4$.}
\label{fig:penrose}
\end{figure}
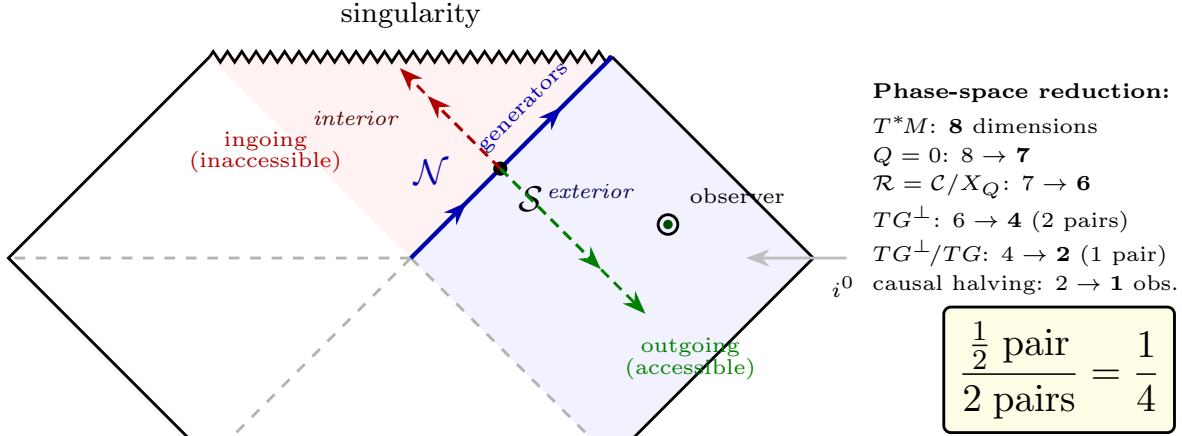

To make contact with the black hole literature, we now
map each step of the general construction onto a
Schwarzschild horizon in ingoing Eddington--Finkelstein
coordinates
\begin{equation}
\dd s^2 = -\!\left(1-\frac{r_s}{r}\right)\dd v^2
  + 2\,\dd v\,\dd r
  + r^2\,\dd\Omega^2\,,
\end{equation}
where $v$ is the advanced time and the metric is
regular at $r=r_s$.
The argument applies to any null boundary, but the
Schwarzschild case provides a concrete identification
of every reduction step (Fig.~\ref{fig:penrose}).

\medskip
\noindent\textbf{Phase space.}
The cotangent bundle $T^*M$ has eight dimensions:
$(v,r,\theta,\varphi,p_v,p_r,p_\theta,p_\varphi)$.

\medskip
\noindent\textbf{Null constraint ($8\to 7$).}
Imposing $Q = g^{ab}p_a p_b = 0$ restricts the momenta
to the null cone.
This selects the hypersurface $\mathcal{C}\subset T^*M$.

\medskip
\noindent\textbf{Quotient by $X_Q$ ($7\to 6$).}
The flow of $X_Q$ moves a point along a null geodesic
to a different affine parameter value.
Quotienting identifies all parametrizations of the same
geodesic, yielding the space of null rays
$\mathcal{R}$.

\medskip
\noindent\textbf{Horizon as null hypersurface.}
The event horizon $\N$ is the surface $r=r_s$.
In these coordinates the metric is non-degenerate there
and the normal covector $n_a = (\dd r)_a$ is null:
$g^{ab}n_a n_b = g^{rr} = 0$ at $r=r_s$.
Because $n^a$ is both normal and tangent to $\N$, it
defines a preferred direction within the horizon:
the generator direction, which coincides with
$\partial/\partial v$ at $r=r_s$.

\medskip
\noindent\textbf{Foliation and cross-section ($6\to 4$).}
The horizon is foliated by generators---curves at
constant $(\theta,\varphi)$ running in the $v$-direction.
Quotienting $\N$ by its generators yields a 2-sphere
$\Sc$ with coordinates $(\theta,\varphi)$.
In $\mathcal{R}$, this corresponds to the isotropic
submanifold $G$, and its symplectic complement
$TG^\perp$ has dimension $6-2=4$: two canonical pairs.
These are the naive phase-space data of a surface
element of $\Sc$.

\medskip
\noindent\textbf{Dressing time ($4\to 2$).}
The isotropic quotient $TG^\perp/TG$ removes the
generator-labeling directions, consuming one canonical
pair.
The surviving pair corresponds to the expansion $\Theta$
and shear $\sigma_{ab}$ of the null congruence---the
Bondi news data on the horizon
\cite{HopfmullerFreidel2017}.

\medskip
\noindent\textbf{Causal halving ($2\to 1$ observable).}
An observer at $r>r_s$ receives signals carried by
outgoing null rays only.
The ingoing component crosses the horizon and is
causally inaccessible from the exterior.
The full phase space is intact---an infalling observer
encounters nothing special at $r=r_s$---but
the exterior observer has access to only half of the
surviving canonical pair.

\medskip
The observable fraction is $\frac{1}{2}$ of one pair
out of a naive two: $1/4$.
Each step in the general construction of
Sections~3--4 has a direct black hole counterpart,
but none of the steps depends on the Einstein equations.
The coefficient $1/4$ emerges from the null causal
structure of the horizon, not from the dynamics that
produces it.

\section{Discussion}

The coefficient $1/4$ is a consequence of
null causal structure and time orientation in four-dimensional
Lorentzian spacetime.
It is independent of the Einstein equations, of any
gravitational action, of Newton's constant (except through
the Planck length in defining the cell size), and of any
particular spacetime solution.

The derivation is explicitly four-dimensional;
the generalization to arbitrary $D$ is under investigation.
The argument requires three ingredients:
a Lorentzian metric, null geodesics, and a closed 2-surface
that acts as a one-way causal boundary.
These are shared by every framework in which black hole entropy
has been computed.
The universality of the coefficient across semiclassical,
string-theoretic, and loop gravity derivations
is therefore not a coincidence to be explained by each
framework separately, but a geometric fact that precedes
all of them.

\medskip
\noindent\textbf{Epistemic interpretation.}
The entropy in this derivation counts accessible
phase-space volume, not missing information.
No degrees of freedom are traced over, integrated out,
or lost.
The full phase space of the null surface is intact;
the causal boundary restricts which portion an exterior
observer can probe.
The factor $1/4$ is the geometric measure of this
epistemic restriction for a null causal boundary.

\medskip
\noindent\textbf{What this derivation does and does not explain.}
The Planck-area cell count $N = A/\lp^2$ is an input,
not a result, and null geodesics are not identified here
as the microscopic degrees of freedom responsible for
black hole entropy.
This paper derives the fraction $1/4$ but not the
proportionality to area, which requires a UV completion
that discretizes the surface into Planck-scale
cells (see, e.g., \cite{Davidson2019}).
The reduction chain does, however, offer a geometric
explanation for why entropy is naturally associated with
a two-dimensional surface: the successive constraints
reduce all observable phase-space degrees of freedom to
those living on the spacelike cross-section $\Sc$,
providing a kinematic basis for the holographic
association of entropy with area.
The argument is kinematic: whatever microscopic framework
supplies the state count---field modes, spin network states,
or combinatorial surface elements---the null causal geometry
fixes the accessible fraction to $1/4$.
The derivation does not count those degrees of freedom,
explain why black holes radiate, why the radiation is
thermal, or how information is returned.
In this sense, the derivation is theory-agnostic: the
factor $1/4$ is a consequence of null causal geometry
alone, independent of whether the underlying microscopic
theory is quantum field theory in curved spacetime,
string theory, loop quantum gravity, or any other
UV completion.

\medskip
\noindent\textbf{Relation to the information paradox.}
The causal halving step implies that the
phase-space data behind the horizon is inaccessible,
not absent.
An infalling observer encounters the full phase space;
the restriction is perspectival, consistent with
complementarity and with the expectation that unitarity
is preserved globally.

\section*{Conflict of Interest}
The author declares no competing interests.

\section*{Data Access Statement}
This work is purely analytical. No experimental data were generated or analysed.

\section*{Ethics Statement}
This research did not involve human subjects, animal subjects, or ethical approval requirements.

\section*{Funding}
This research received no specific grant from any funding agency.


\section*{References}
\begin{thebibliography}{10}

\bibitem{Bekenstein1973}
J.~D.~Bekenstein,
Phys.\ Rev.\ D \textbf{7}, 2333 (1973).

\bibitem{Hawking1975}
S.~W.~Hawking,
Commun.\ Math.\ Phys.\ \textbf{43}, 199 (1975).

\bibitem{Wald1993}
R.~M.~Wald,
Phys.\ Rev.\ D \textbf{48}, R3427 (1993).

\bibitem{IyerWald1994}
V.~Iyer and R.~M.~Wald,
Phys.\ Rev.\ D \textbf{50}, 846 (1994).

\bibitem{GibbonsHawking1977}
G.~W.~Gibbons and S.~W.~Hawking,
Phys.\ Rev.\ D \textbf{15}, 2752 (1977).

\bibitem{Jacobson1995}
T.~Jacobson,
Phys.\ Rev.\ Lett.\ \textbf{75}, 1260 (1995).

\bibitem{Davidson2019}
A.~Davidson,
Phys.\ Rev.\ D \textbf{100}, 081502(R) (2019).

\bibitem{DonnellyFreidel2016}
W.~Donnelly and L.~Freidel,
JHEP \textbf{09}, 102 (2016).

\bibitem{HopfmullerFreidel2017}
F.~Hopfm\"uller and L.~Freidel,
Phys.\ Rev.\ D \textbf{95}, 104006 (2017).

\bibitem{Rovelli2002}
C.~Rovelli,
Phys.\ Rev.\ D \textbf{65}, 124013 (2002).

\bibitem{GambiniPullin2011}
R.~Gambini and J.~Pullin,
\textit{A First Course in Loop Quantum Gravity}
(Oxford University Press, 2011).

\bibitem{RovelliVidotto2015}
C.~Rovelli and F.~Vidotto,
\textit{Covariant Loop Quantum Gravity}
(Cambridge University Press, 2015), Chap.~2.

\bibitem{CiambelliFreidelLeigh2024}
L.~Ciambelli, L.~Freidel, and R.~G.~Leigh,
JHEP \textbf{01}, 166 (2024).

\bibitem{WaldGR1984}
R.~M.~Wald,
\textit{General Relativity} (University of Chicago Press, 1984).

\bibitem{AbrahamMarsden1978}
R.~Abraham and J.~E.~Marsden,
\textit{Foundations of Mechanics},
2nd ed.\ (Benjamin/Cummings, 1978).

\bibitem{Arnold1989}
V.~I.~Arnold,
\textit{Mathematical Methods of Classical Mechanics},
2nd ed.\ (Springer, 1989).

\bibitem{Speranza2018}
A.~J.~Speranza,
JHEP \textbf{02}, 021 (2018).

\bibitem{ChandrasekaranSperanza2021}
V.~Chandrasekaran and A.~J.~Speranza,
JHEP \textbf{01}, 137 (2021).

\bibitem{Raychaudhuri1955}
A.~K.~Raychaudhuri,
Phys.\ Rev.\ \textbf{98}, 1123 (1955).

\end{thebibliography}
\end{document}